\documentstyle[twoside,fleqn,espcrc2,epsfig]{article}

\def\defeq{\mathrel{\mathop=^{\rm def}}}
\def\proof{\noindent{\sl Proof:}\kern0.6em}

\def\frac#1#2{\hbox{$#1\over#2$}}
\def\dual{\mathstrut^*\kern-0.1em}

\def\lvec#1{\setbox0=\hbox{$#1$}
    \setbox1=\hbox{$\scriptstyle\leftarrow$}
    #1\kern-\wd0\smash{
    \raise\ht0\hbox{$\raise1pt\hbox{$\scriptstyle\leftarrow$}$}}
    \kern-\wd1\kern\wd0}
\def\rvec#1{\setbox0=\hbox{$#1$}
    \setbox1=\hbox{$\scriptstyle\rightarrow$}
    #1\kern-\wd0\smash{
    \raise\ht0\hbox{$\raise1pt\hbox{$\scriptstyle\rightarrow$}$}}
    \kern-\wd1\kern\wd0}

\def\nabstar#1{\nabla\kern-0.5pt\smash{\raise 4.5pt\hbox{$\ast$}}
               \kern-4.5pt_{#1}}

\def\drvstar#1{\partial\kern-0.5pt\smash{\raise 4.5pt\hbox{$\ast$}}
               \kern-5.0pt_{#1}}

\def\Nf{N_{\rm f}}
\def\psibar{\overline{\psi}}

\def\rhoprime{\rho\kern1pt'}
\def\rhobar{\bar{\rho}}
\def\rhobarprime{\rhobar\kern1pt'}
\def\rhobartilde{\kern2pt\tilde{\kern-2pt\rhobar}}
\def\rhobartildeprime{\kern2pt\tilde{\kern-2pt\rhobar}\kern1pt'}

\def\zetabar{\bar{\zeta}}
\def\zetaprime{\zeta\kern1pt'}
\def\zetabarprime{\zetabar\kern1pt'}
\def\zetar{\zeta_{\raise-1pt\hbox{\sixrm R}}}
\def\zetabarr{\zetabar_{\raise-1pt\hbox{\sixrm R}}}

\def\phiimpr{\phi_{\kern0.5pt\hbox{\sixrm I}}}

\def\ar{A_{\mbox{\scriptsize{\rm R}}}}
\def\vr{V_{\mbox{\scriptsize{\rm R}}}}
\def\pr{P_{\mbox{\scriptsize{\rm R}}}}

\def\diracstar#1#2{
    \setbox0=\hbox{$\gamma$}\setbox1=\hbox{$\gamma_{#1}$}
    \gamma_{#1}\kern-\wd1\kern\wd0
    \smash{\raise4.5pt\hbox{$\scriptstyle#2$}}}

\def\ba{b_{\rm A}}
\def\bv{b_{\rm V}}
\def\bp{b_{\rm P}}

\def\bg{b_{\rm g}}
\def\bm{b_{\rm m}}

\def\ca{c_{\rm A}}
\def\cv{c_{\rm V}}

\def\f1{f_1}

\def\SUtwo{{\rm SU(2)}}

\def\tr{\,\hbox{tr}\,}

\def\opprime#1{\setbox0=\hbox{${\cal O}$}\setbox1=\hbox{${\cal O}_{\rm #1}$}
    {\cal O}_{\rm #1}\kern-\wd1\kern\wd0
    \smash{\raise4.5pt\hbox{\kern1pt$\scriptstyle\prime$}}\kern1pt}

\def\ophatprime#1{\setbox0=\hbox{$\widehat{\cal O}$}
    \setbox1=\hbox{$\widehat{\cal O}_{\rm #1}$}
    \widehat{\cal O}_{\rm #1}\kern-\wd1\kern\wd0
    \smash{\raise4.5pt\hbox{\kern1pt$\scriptstyle\prime$}}\kern1pt}

\def\bopprime#1{\setbox0=\hbox{${\cal O}$}\setbox1=\hbox{${\cal O}_{\rm #1}$}
    {\cal L}_{\rm #1}\kern-\wd1\kern\wd0
    \smash{\raise4.5pt\hbox{\kern1pt$\scriptstyle\prime$}}\kern1pt}

\def\blagprime#1{\setbox0=\hbox{${\cal B}$}\setbox1=\hbox{${\cal B}_{#1}$}
    {\cal B}_{#1}\kern-\wd1\kern\wd0
    \smash{\raise5.2pt\hbox{\kern1pt$\scriptstyle\prime$}}\kern1pt}

\def\mq{m_{\rm q}}
\def\mqtilde{\widetilde{m}_{\rm q}}
\def\mr{m_{{\mbox{\scriptsize{\rm R}}}}}
\def\mc{m_{\rm c}}

\def\za{Z_{\rm A}}
\def\zp{Z_{\rm P}}
\def\zv{Z_{\rm V}}
\def\zg{Z_{\rm g}}
\def\zm{Z_{\rm m}}

\def\gtilde{\tilde{g}_0}

\def\msbar{{\rm \overline{MS\kern-0.05em}\kern0.05em}}

\def\defeq{\mathrel{\mathop=^{\rm def}}}
\def\proof{\noindent{\sl Proof:}\kern0.6em}

\def\frac#1#2{\hbox{$#1\over#2$}}
\def\dual{\mathstrut^*\kern-0.1em}

\def\lvec#1{\setbox0=\hbox{$#1$}
    \setbox1=\hbox{$\scriptstyle\leftarrow$}
    #1\kern-\wd0\smash{
    \raise\ht0\hbox{$\raise1pt\hbox{$\scriptstyle\leftarrow$}$}}
    \kern-\wd1\kern\wd0}
\def\rvec#1{\setbox0=\hbox{$#1$}
    \setbox1=\hbox{$\scriptstyle\rightarrow$}
    #1\kern-\wd0\smash{
    \raise\ht0\hbox{$\raise1pt\hbox{$\scriptstyle\rightarrow$}$}}
    \kern-\wd1\kern\wd0}

\def\nabstar#1{\nabla\kern-0.5pt\smash{\raise 4.5pt\hbox{$\ast$}}
               \kern-4.5pt_{#1}}

\def\drvstar#1{\partial\kern-0.5pt\smash{\raise 4.5pt\hbox{$\ast$}}
               \kern-5.0pt_{#1}}

\def\momp#1#2{
    \setbox0=\hbox{${#1}$}\setbox1=\hbox{${#1}_{#2}$}
    {#1}_{#2}\kern-\wd1\kern\wd0
    \smash{\raise4.5pt\hbox{$\scriptscriptstyle +$}}}
\def\momm#1#2{
    \setbox0=\hbox{${#1}$}\setbox1=\hbox{${#1}_{#2}$}
    {#1}_{#2}\kern-\wd1\kern\wd0
    \smash{\raise4.5pt\hbox{$\scriptscriptstyle -$}}}
\def\mompm#1#2{
    \setbox0=\hbox{${#1}$}\setbox1=\hbox{${#1}_{#2}$}
    {#1}_{#2}\kern-\wd1\kern\wd0
    \smash{\raise4.5pt\hbox{$\scriptscriptstyle \pm$}}}
\def\smomp#1#2{
    \setbox0=\hbox{${#1}$}\setbox1=\hbox{${#1}_{#2}$}
    {#1}_{#2}\kern-\wd1\kern\wd0
    \smash{\raise3pt\hbox{$\scriptscriptstyle +$}}}
\def\smomm#1#2{
    \setbox0=\hbox{${#1}$}\setbox1=\hbox{${#1}_{#2}$}
    {#1}_{#2}\kern-\wd1\kern\wd0
    \smash{\raise3pt\hbox{$\scriptscriptstyle -$}}}
\def\smompm#1#2{
    \setbox0=\hbox{${#1}$}\setbox1=\hbox{${#1}_{#2}$}
    {#1}_{#2}\kern-\wd1\kern\wd0
    \smash{\raise3pt\hbox{$\scriptscriptstyle \pm$}}}

\def\si{\kern1pt{\rm si}}
\def\co{\kern1pt{\rm co}}

\def\Nf{N_{\rm f}}
\def\psibar{\overline{\psi}}
\def\psiL{\psi_{\rm L}}
\def\psiR{\psi_{\rm R}}
\def\psibarL{\psibar_{\rm L}}
\def\psibarR{\psibar_{\rm R}}

\def\rhoprime{\rho\kern1pt'}
\def\rhobar{\bar{\rho}}
\def\rhobarprime{\rhobar\kern1pt'}
\def\rhobartilde{\kern2pt\tilde{\kern-2pt\rhobar}}
\def\rhobartildeprime{\kern2pt\tilde{\kern-2pt\rhobar}\kern1pt'}

\def\zetabar{\bar{\zeta}}
\def\zetaprime{\zeta\kern1pt'}
\def\zetabarprime{\zetabar\kern1pt'}
\def\zetar{\zeta_{\raise-1pt\hbox{\sixrm R}}}
\def\zetabarr{\zetabar_{\raise-1pt\hbox{\sixrm R}}}

\def\phiimpr{\phi_{\kern0.5pt\hbox{\sixrm I}}}

\def\ar{A_{\hbox{\sixrm R}}}
\def\vr{V_{\hbox{\sixrm R}}}
\def\pr{P_{\hbox{\sixrm R}}}

\def\aimp{A_{\hbox{\sixrm I}}}
\def\vimp{V_{\hbox{\sixrm I}}}
\def\pimp{P_{\hbox{\sixrm I}}}

\def\diracstar#1#2{
    \setbox0=\hbox{$\gamma$}\setbox1=\hbox{$\gamma_{#1}$}
    \gamma_{#1}\kern-\wd1\kern\wd0
    \smash{\raise4.5pt\hbox{$\scriptstyle#2$}}}

\def\ba{b_{\rm A}}
\def\tba{\tilde{b}_{\rm A}}
\def\bp{b_{\rm P}}

\def\bv{b_{\rm V}}
\def\tbv{\tilde{b}_{\rm V}}

\def\bg{b_{\rm g}}
\def\bm{b_{\rm m}}
\def\tbm{\tilde{b}_{\rm m}}
\def\bmu{b_{\mu}}

\def\ca{c_{\rm A}}
\def\cv{c_{\rm V}}

\def\f1{f_1}

\def\h1{h_1}

\def\SUtwo{{\rm SU(2)}}

\def\tr{\,\hbox{tr}\,}

\def\opprime#1{\setbox0=\hbox{${\cal O}$}\setbox1=\hbox{${\cal O}_{\rm #1}$}
    {\cal O}_{\rm #1}\kern-\wd1\kern\wd0
    \smash{\raise4.5pt\hbox{\kern1pt$\scriptstyle\prime$}}\kern1pt}

\def\ophatprime#1{\setbox0=\hbox{$\widehat{\cal O}$}
    \setbox1=\hbox{$\widehat{\cal O}_{\rm #1}$}
    \widehat{\cal O}_{\rm #1}\kern-\wd1\kern\wd0
    \smash{\raise4.5pt\hbox{\kern1pt$\scriptstyle\prime$}}\kern1pt}

\def\bopprime#1{\setbox0=\hbox{${\cal O}$}\setbox1=\hbox{${\cal O}_{\rm #1}$}
    {\cal L}_{\rm #1}\kern-\wd1\kern\wd0
    \smash{\raise4.5pt\hbox{\kern1pt$\scriptstyle\prime$}}\kern1pt}

\def\blagprime#1{\setbox0=\hbox{${\cal B}$}\setbox1=\hbox{${\cal B}_{#1}$}
    {\cal B}_{#1}\kern-\wd1\kern\wd0
    \smash{\raise5.2pt\hbox{\kern1pt$\scriptstyle\prime$}}\kern1pt}

\def\muq{\mu_{\rm q}}
\def\mq{m_{\rm q}}
\def\mqtilde{\widetilde{m}_{\rm q}}
\def\muqtilde{\widetilde{\mu}_{\rm q}}

\def\mr{m_{{\hbox{\sixrm R}}}}
\def\mur{\mu_{{\hbox{\sixrm R}}}}
\def\mc{m_{\rm c}}

\def\za{Z_{\rm A}}
\def\zv{Z_{\rm V}}
\def\zp{Z_{\rm P}}

\def\zg{Z_{\rm g}}
\def\zm{Z_{\rm m}}

\def\Za{\za}
\def\Zv{\zv}
\def\Zp{\zp}

\def\Zg{\zg}
\def\Zm{\zm}

\def\gtilde{\tilde{g}_0}

\def\msbar{{\rm \overline{MS\kern-0.05em}\kern0.05em}}

\title{A local formulation of lattice QCD without unphysical fermion
zero modes\thanks{based on a talk by R.~Frezzotti and a poster by S.~Sint,
presented at the International Symposium on 
Lattice Field Theory, June 29 -- July 3, 1999, Pisa, Italy}}

\author{Roberto Frezzotti$^{\rm a}$, 
Pietro Antonio Grassi\address{Max-Planck-Institut f\"ur Physik, F\"ohringer 
                      Ring 6, D-80805 M\"unchen, Germany},
Stefan Sint\address{Universit\`a di Roma ``Tor
Vergata'', Dipartimento di Fisica, Via della Ricerca Scientifica
                      1, I-00133 Roma, Italy}
and Peter Weisz$^{\rm a}$
}

\begin{document}
\begin{abstract}
The  problem of unphysical zero modes in lattice QCD with Wilson 
fermions can be solved in a clean way
by including a mass term proportional to $i \psibar \gamma_5 \tau^3 \psi$
in the standard lattice theory with $\Nf=2$ mass degenerate Wilson quarks. 
We argue  that up to cutoff effects, this lattice theory 
is equivalent to standard lattice QCD, for suitable choices
of the mass parameters and with a natural re-interpretation of observables. 
On-shell O($a$) improvement can be implemented in a straightforward way.
\end{abstract}
\maketitle

\section{INTRODUCTION}

Lattice QCD with Wilson quarks provides an attractive framework
for non-perturbative computations of many
interesting phenomenological quantities. Its main disadvantage 
consists in the explicit breaking of all axial symmetries.
As a consequence the Wilson-Dirac operator 
is not protected against zero modes even at finite values 
of the quark mass. While this is not a theoretical 
problem for the full theory,
it severely limits the applicability of the quenched approximation as 
the occurrence of unusually small eigenvalues
may lead to large fluctuations in observable quantities. 
In practice this means that one cannot directly treat 
the light quark masses but has to extrapolate 
results obtained with relatively heavy quarks.
It is important to appreciate that this limitation is completely 
independent of the usual requirement of not too light pion masses,
dictated by the finite space-time volume used for the simulation.
Indeed quite often quenched simulations are limited by the occurrence of 
``exceptional configurations" at quark mass values where
finite volume effects are still completely negligible.

Various recipes have been proposed to deal with this 
problem~[1--3] but to our knowledge 
no completely satisfactory solution has appeared in the literature. 
In order to fill this gap, we consider the lattice action
first introduced by Aoki~\cite{Aoki}.
More precisely, if $D_{\rm W}$ denotes the
(O($a$) improved) Wilson-Dirac operator for a quark with
bare mass $m_0$, we consider the 
fermionic action for $\Nf=2$ quarks with the lattice Dirac operator 
\begin{equation}
  D_{\rm twist}\defeq D_{\rm W} +i\muq\gamma_5\tau^3.
\end{equation}
The additional term has a non-trivial flavour structure 
(the Pauli matrix $\tau^3$ acts in flavour space), and
will be referred to as a chirally twisted mass term.
It is obvious that this lattice Dirac operator 
is protected against zero modes for any finite value of $\muq$,
\begin{equation}
  \det D_{\rm twist} = \det( D_{\rm W}^\dagger  D_{\rm W}^{} +\muq^2) >0. 
\end{equation}
To use this lattice action as a regularization of 
``exceptional configurations'' has therefore 
been proposed in ref.~\cite{BardeenII}, 
implying, however, that the mass parameter $\muq$ be 
extrapolated to zero at the end.
As will become clear below, this is not necessary. 
It can be shown that lattice QCD with a chirally
twisted mass term (twisted lattice QCD) is,
up to cutoff effects, completely equivalent to standard lattice QCD,
and thus provides an alternative regularization of the massive theory. 

This article is organised as follows:
we first take a closer look at the classical continuum
limit of twisted lattice QCD (sect.~2). Assuming universality 
we sketch an elementary proof that twisted and standard lattice QCD 
are equivalent in the continuum limit (sect.~3). 
In sect.~4 we determine the possible counterterms,
and briefly introduce our choice of renormalization scheme.
We then discuss O($a$) improvement (sect.~5) 
and we end with a few comments on further applications.

\section{CLASSICAL CONTINUUM THEORY}

To get acquainted with twisted QCD
it is instructive to have a closer look at the 
corresponding classical continuum theory, 
described by the lagrangian
\begin{equation}
{\cal L}_f(x) =\psibar(x)\left(D\kern-7pt\slash
                +m+i\muq\gamma_5\tau^3\right)\psi(x). 
\end{equation}
Defining the usual isospin currents and densities,
\begin{eqnarray}
  A_\mu^a &=& \psibar\gamma_\mu\gamma_5{{\tau^a}\over{2}}\psi,\\
  V_\mu^a &=& \psibar\gamma_\mu{{\tau^a}\over{2}}\psi,\\
  P^a     &=& \psibar\gamma_5{{\tau^a}\over{2}}\psi,
\end{eqnarray}
the partial conservation of axial and vector currents reads
\footnote{We define the antisymmetric
tensor $\varepsilon_{ab}$ with indices 
$a,b=1,2$ and normalization $\varepsilon_{12}=1$.
Summation over repeated indices is always understood.}
\begin{eqnarray}
   \partial_\mu A_\mu^a &=& 2m P^a, \quad a=1,2\\
   \partial_\mu A_\mu^3 &=& 2m P^3+i\muq\psibar\psi,\\
   \partial_\mu V_\mu^a &=& -2\muq\,\varepsilon_{ab} P^b, \quad a,b=1,2\\
   \partial_\mu V_\mu^3 &=& 0. 
\end{eqnarray}
A global chiral (non-singlet) rotation of the fields,
\begin{eqnarray}
 \psi'    &=&\exp(i \alpha\gamma_5\tau^3/2)\psi,\nonumber\\
 \psibar' &=&\psibar\exp(i \alpha\gamma_5\tau^3/2),\label{axialI}
\end{eqnarray}
with  $\tan\alpha = \muq/\mq$
transforms the Lagrangian to its standard form,
\begin{equation}
  {\cal L}'_f(x) =\psibar'(x)\left(D\kern-7pt\slash
                +m'\right)\psi'(x), 
\end{equation}
with the  quark mass $m'= \sqrt{m^2+\muq^2}\,$.
The isospin currents and densities in the primed basis are
related to the original fields through ($a,b=1,2$) 
\begin{eqnarray}
  {A'}_\mu^a &=& \cos(\alpha) A_\mu^a + \sin(\alpha) 
                 \varepsilon_{ab} V_\mu^b,\\  
  {A'}_\mu^3 &=& A_\mu^3,\\
  {V'}_\mu^a &=& \cos(\alpha) V_\mu^a + \sin(\alpha) 
                 \varepsilon_{ab} A_\mu^b,\\  
  {V'}_\mu^3 &=& V _\mu^3,\\ 
  {P'}^a     &=& P^a,\\
  {P'}^3     &=& \cos(\alpha) P^3 + i\sin(\alpha) \frac{1}{2}\psibar\psi,
\end{eqnarray}
and satisfy the PCAC and PCVC relations in their standard forms,
i.e.
\begin{equation}
   \partial_\mu {A'}_\mu^a = 2m'{P'}^a, \quad \partial_\mu {V'}_\mu^a = 0, \quad a=1,2,3.
\end{equation}
Similar relationships can be worked out between more complicated
composite fields
, and we thus conclude that 
the classical twisted theory can be mapped back
to the standard formulation by a simple axial rotation of the fields, with 
a unique rotation angle which is fixed by the ratio of the
mass parameters. Both formulations are
thus equivalent and contain the same physical information.

\section{BEYOND THE CLASSICAL THEORY}

In order to establish the equivalence 
between twisted QCD and the standard formulation of QCD
beyond the classical continuum limit,
it is convenient to make use of a regularization 
which does not break the flavour symmetries.
This is possible~\cite{Hasenfratz,LuscherI} 
if the lattice Dirac operator satisfies the
Ginsparg-Wilson relation~\cite{GW},
\begin{equation}
  D \gamma_5+\gamma_5 D = a D \gamma_5 D.
\end{equation}
An explicit solution for $D$ 
has been given by Neuberger~\cite{Neuberger},
but in the following we only need to know that
$D$ is a local operator which satisfies the Ginsparg-Wilson
relation and has the conjugation property $D^\dagger=\gamma_5 D\gamma_5$.
It then follows that the matrix~\cite{LuscherII}, 
\begin{equation}
  \hat\gamma_5\defeq \gamma_5(1-aD),
\end{equation}
is hermitian and unitary.
Therefore it may be used to define
left handed quark fields through
\begin{equation}
 \psiL = \frac12(1-\hat\gamma_5)\psi, \qquad
 \psibarL = \psibar\frac12(1+\gamma_5),
\end{equation}
and right handed fields are obtained with the complementary 
projectors.
Using these chiral fields a lattice regularized version of
twisted QCD is specified by the action 
\begin{eqnarray}
  S_f &=& a^4\sum_{x}\Bigl[ \psibarL D \psiL 
        + \psibarR D \psiR \nonumber\\
     &&\hphantom{a^4\sum_{x}\Bigl[}
         + m(\psibarL \psiR + \psibarR \psiL) \nonumber\\
     &&\hphantom{a^4\sum_{x}} 
         + i\muq(\psibarL\tau^3 \psiR - \psibarR \tau^3 \psiL)\Bigr].  
\end{eqnarray} 
In the chiral limit $m=\muq=0$ this action has an exact
$\SUtwo\times\SUtwo$ flavour symmetry, and in particular
the axial transformation [cf.~eq.~(\ref{axialI})],
\begin{eqnarray}
  \psiL'&=&\exp(-i\alpha\tau^3/2)\,\psiL,\nonumber\\
  \psibarL'&=&\psibarL\exp(i\alpha\tau^3/2),\nonumber\\
  \psiR'&=&\exp(i\alpha\tau^3/2)\,\psiR,\nonumber\\
  \psibarR'&=&\psibarR\exp(-i\alpha\tau^3/2) \label{axialII},
\end{eqnarray}
is part of the symmetry group.
Therefore, this transformation leaves the form of 
the lattice action invariant and merely
transforms the  parameters, 
\begin{eqnarray}
  m'&=&m\cos(\alpha)+\muq\sin(\alpha),\\
  \muq'&=&-m\sin(\alpha)+\muq\cos(\alpha),
\end{eqnarray}
where the choice  $\tan\alpha=\muq/m$
again produces the standard action with $\muq'=0$ and
$m'=(m^2+\muq^2)^{1/2}$.
Since the flavour symmetry in the 
basis of chiral fields takes exactly 
the same form as in the continuum, we 
may also reproduce eqs.~(13--19). More generally,
bare operators in a given representation of the 
flavour symmetry group are rotated covariantly
by the same angle $\alpha$. 
Finally we note that the measure of the functional integral is left
invariant under the axial (non-singlet) transformation, eq.~(\ref{axialII}). 
We have thus found a regularization where (at least in a finite 
space-time volume) the bare parameters and correlation functions 
of twisted QCD are related to the ones of standard QCD 
through a change of variables in a well-defined functional integral.
This yields exact identities at any fixed value of the lattice spacing,
and gives a precise meaning to the equivalence between the two
theories. For example, if $O_k^{}$  and $O_k'$, are, for $k=1,\ldots,n$, 
bare gauge invariant composite operators, which are related through the
axial transformation~(\ref{axialII}), we obtain the identity
between the (bare) connected correlation functions,
\begin{eqnarray}
 && \langle O_1(x_1)\cdots O_n(x_n) \rangle_c^{\rm twisted\,QCD} =\nonumber\\ 
 &&\langle O_1'(x_1) \cdots O_n'(x_n)  \rangle_c,
\end{eqnarray}
provided the bare parameters $\muq,m$ and $m'$
are related as indicated previously.


In order for these relations to carry over to the 
continuum limit, it is convenient to  
renormalize both mass parameters, $\muq$ and $m$ 
in the same way, such that 
the ratio $\muq/m$, and thus the angle $\alpha$, 
remains unchanged.
If the chosen renormalization scheme respects the global flavour
symmetries, then the relations between correlation functions of 
composite operators are exactly the same as in the bare theory,
and thus also hold in the continuum limit. 
Mass independent renormalization schemes obviously have this 
property and in addition preserve relations between renormalized
parameters, but it turns out that a much wider class of
renormalization schemes is admissible~\cite{workinprogress}.

In lattice QCD with Wilson fermions an axial transformation
does not correspond to an exact symmetry. However, 
assuming universality of the continuum limit,
we may conclude that the twisted and standard formulations 
of lattice QCD with Wilson fermions are indeed 
equivalent up to cutoff effects. 
In fact, this can be proven rigorously to all orders in the 
loop expansion, by exploiting reparametrization invariance and the
Quantum Action Principle~\cite{workinprogress}. 
In particular, this proof
holds for any regularization for which a power counting theorem 
can be established, regardless of whether or not the 
flavour symmetries are broken by the regularization.

\section{RENORMALIZATION}

On the lattice with Wilson fermions only the isospin $\SUtwo$
symmetry remains exact, and the flavour symmetry is
further reduced by the introduction of the twisted mass term. 
This is in contrast with the lattice regularization based on
Ginsparg-Wilson fermions, where the symmetries are exactly the same 
in both (standard and twisted) formulations, 
albeit realized in a different way.

A careful analysis shows that, apart from gauge invariance
and invariance under discrete four-dimensional rotations,
the following symmetries remain exact on the lattice:
\begin{itemize}
\item a global U(1) symmetry corresponding to fermion number conservation,
\item a residual U(1) flavour symmetry with generator $\tau^3/2$,
\item charge conjugation,
\item $P_F$: parity combined with flavour exchange,
\item $\tilde{F}$: flavour exchange combined with $\muq \rightarrow -\muq$.
\end{itemize}
Furthermore, it is possible to 
show that reflection positivity \cite{refl_pos} still holds.

Taking into account these symmetries, twisted
lattice QCD is renormalizable by power counting. The
only new counterterm of dimension $\leq 4$ 
is proportional to  $i\muq\psibar\gamma_5\tau^3\psi$, which
leads to a multiplicative renormalization of $\muq$.
In particular we note that a term of the form
$\varepsilon_{\mu\nu\rho\sigma}\tr\{F_{\mu\nu}F_{\rho\sigma}\}$
is excluded by the symmetry $P_F$, whereas invariance under
$\tilde{F}$ excludes an additive renormalization
of $\muq$ 
(i.e.~a counterterm proportional to $a^{-1}i\psibar\gamma_5\tau^3\psi$).


We would like to impose renormalization conditions such that the
relation between renormalized lattice QCD and its
twisted version is as simple as in the classical continuum limit. 
However, due to the reduced flavour 
symmetry with Wilson fermions some preparation is
needed. In particular, one has to determine 
the relative normalization between composite operators that belong 
to the same representation of the continuum flavour symmetry,
but are not related by lattice symmetries.
This can be achieved  by imposing continuum 
chiral Ward identities as normalization conditions
for the lattice operators~\cite{BochicchioEtAl}. 
An example is provided by the isospin currents: 
while the conserved vector current is protected 
against renormalization, the correctly normalized
axial current $(\ar)_\mu^a$ has to be 
determined by imposing current algebra relations
as normalization conditions~\cite{paperIV}. We then expect that relations
like eqs.(13--18) hold for the correctly normalized currents, provided the
angle $\alpha$ is defined using the Ward identity masses. 
More precisely, we define a bare current quark mass $m$
through (some suitable matrix element of) the PCAC relation, 
\begin{equation}
   \partial_\mu (\ar)_\mu^1 = 2 m P^1,
\end{equation}
and define $\alpha$ through
\begin{equation}
  \tan\alpha = \muq/m.
\end{equation}
Note that $m$ differs from the subtracted bare
quark mass $\mq=m_0-\mc$. 
In the renormalized theory, we thus fix
the ratio of the mass renormalization constants
\begin{equation}
  \mr = \Zm \mq, \qquad \mur = Z_\mu\muq,
\end{equation}
such that it coincides with the slope at
the origin of the function $m(\mq)$.

\section{O($a$) IMPROVEMENT}

Given the symmetries of the lattice theory it
is a straightforward exercise to list the
possible O($a$) counterterms to the action. 
The procedure has been explained in great detail
in ref.~\cite{paperI} and applies to the case
at hand, taking into account the modified 
set of symmetries given in the previous section.
We expect the renormalization and O($a$) improvement 
of the massless theory to be the same as in standard
lattice QCD~[13--15].
This follows from
analyticity of correlation functions
in the mass parameters $\mq$ and $\muq$, which is
however guaranteed only with an infrared cutoff, e.g.
in a finite space-time volume.

With this assumption, on-shell O($a$) improvement 
of the action is  achieved by using the Sheikholeslami-Wohlert improved
Wilson-Dirac operator~\cite{SW}, 
and the following parameterization of the
O($a$) improved renormalized coupling and quark masses,
\begin{eqnarray}
  g_{\rm R}^2 &=&  \gtilde^2\Zg(\gtilde^2,a\mu),\\
  m_{\rm R}   &=&  \mqtilde\Zm(\gtilde^2,a\mu),\\
  \mu_{\rm R} &=&  \muqtilde Z_\mu(\gtilde^2,a\mu),\\
  \gtilde^2   &=& g_0^2(1+\bg a\mq),\\
  \mqtilde    &=& \mq + b_{\rm m} a\mq^2 +\tbm a\muq^2,\\
  \muqtilde   &=& \muq(1+\bmu a\mq),\\
  \mq         &=& m_0-\mc.
\end{eqnarray}
This notation already implies our choice of a mass-independent
renormalization scheme~\cite{Weinberg,letter}.
O($a$) improvement of the action thus 
introduces the improvement coefficients $\bmu$ and $\tbm$,
in addition to the standard coefficients $\bm$ and $\bg$.

As for the improvement of operators we here give a few examples
which appear in the PCAC and PCVC relations.
Following ref.~\cite{paperI} we define the improved operators
($a=1,2$),
\begin{eqnarray}
  (\aimp)_\mu^a &=& A_\mu^a+\ca a\tilde\partial_\mu
                         P^a + a\muq\tba \varepsilon_{ab} V_\mu^b, \\
  (\vimp)_\mu^a &=& V_\mu^a+\cv a\tilde\partial_\nu
                 T^a_{\mu\nu}+ a\muq\tbv \varepsilon_{ab} A_\mu^b,\\
  (\pimp)^a &=& P^a,
\end{eqnarray}
and obtain the renormalized improved operators 
through ($a=1,2$):
\begin{eqnarray}
  (\ar)_\mu^a &=& \Za(1+\ba a\mq)(\aimp)_\mu^a, \\
  (\vr)_\mu^a &=& \Zv(1+\bv a\mq)(\vimp)_\mu^a, \\
  (\pr)^a &=& \Zp(1+\bp a\mq)(\pimp)^a.
\end{eqnarray}
Note that we may avoid technical problems associated with power divergences by
restricting  attention to the first two isospin components of these operators.
This can be done without any loss of information due to the expected restoration
of all chiral Ward identities in the continuum limit.

With these definitions the O($a$) improved angle $\alpha$
is defined by
\begin{equation}
 \tan\alpha = {{\mur}\over{\mr}} = {{Z_\mu \muqtilde}\over{\Zm \mqtilde}} 
\end{equation}
where $\mur$ and $\mr$ are obtained from matrix elements of the
renormalized O($a$) improved PCAC and PCVC relations.
Again this requires that the ratio $Z_\mu/\Zm$ be fixed as explained in the
previous section.

In order to check the general framework of renormalization and
O($a$) improvement described above, we carried out
a perturbative calculation using on-shell correlation functions
derived from the  Schr\"odinger functional~\cite{Stefan}. The details
of the calculations are very similar to the studies of
the standard theory \cite{paperII,PeterStefan}
and will be described elsewhere~\cite{workinprogress}.
As a result we confirm the theoretical expectations.
In particular, all improvement coefficients are found to be 
functions of the bare coupling alone, and do not depend e.g.~on
the angle $\alpha$. To lowest order we find 
$\tba^{(0)}=\tbv^{(0)}=\bmu^{(0)}=0$ and $ \tilde{b}_m^{(0)}=-\frac12$,
and the corresponding one-loop coefficients 
are available, too~\cite{workinprogress}.

\section{CONCLUSIONS AND OUTLOOK}

There remains little doubt that twisted lattice QCD 
and standard lattice QCD for $\Nf=2$ mass degenerate
quarks are equivalent theories in the continuum limit,
up to O($a^2$) effects if O($a$) improvement is implemented.
Since twisted lattice QCD is protected against unphysical
fermion zero modes, the problem of exceptional configurations
has thus found a clean solution. 
Furthermore, realistic studies of QCD with $\Nf>2$
can be easily done by using the (improved) standard lattice action 
for the heavier quarks.

In full QCD the problem of exceptional configurations
is absent in principle. However, it has been argued that 
close to the chiral limit, standard algorithms 
may still experience technical problems, showing up 
e.g.~as a poor sampling of configuration space~\cite{KarlRainer,PHMC}.
Twisted QCD may therefore be useful in this
context, too. Incidentally, full QCD simulations
using (unimproved) twisted lattice QCD have already been carried
out, albeit with a different motivation~\cite{Bitar}.

In view of O($a$) improvement, there are preferred choices of $\alpha$. 
If $\alpha$ is small, perturbative estimates 
of the new $b$-coefficients are sufficient, 
while large fluctuations due to close zero modes are
already excluded. Another attractive possibility is the choice
$\alpha=\pi/2$, such that the physical quark mass is
entirely defined in terms of $\mur$. In this case
all the standard $b$-coefficients are not needed, and, in particular,
$\gtilde=g_0$. This latter fact may be interesting for 
full QCD simulations, as continuum and 
chiral extrapolations may become easier.

Since theories with  different values of $\alpha$ are
not related by a lattice symmetry, operator mixing problems 
may look quite different. A simple example is
provided by the iso-singlet scalar density where the usual 
additive renormalization at $\alpha=0$ can be avoided
by working at $\alpha=\pi/2$. 

Dynamical fermion simulations being very expensive 
it has been proposed to obtain information about full QCD
by extrapolating from the unphysical region 
where $\Nf\leq 0$~[26--28]. 
For this ``bermion'' approach the problem of close zero modes 
is much more severe than in the quenched approximation, and may
again be solved using twisted lattice QCD.

Originally, Aoki introduced the chirally twisted mass term to discuss 
the phase structure of the standard lattice theory~\cite{Aoki}. 
At strong coupling he established the existence of a
phase with spontaneous breaking of parity.
We do not exclude that such an Aoki phase also exists at weaker couplings.
However, even if it does, this would be relevant only in large space-time 
volume and very close to the chiral limit 
(mass parameters of O($a^2$)) where simulations are not practical
anyway (cf.~ref.~\cite{SharpeSingleton}).

In conclusion, we believe that twisted lattice QCD provides
a very promising alternative formulation of lattice QCD.

\vskip 1ex

This work is part of the ALPHA collaboration research programme.
We are grateful to M.~L\"uscher for useful suggestions,
and to R.~Sommer for a critical reading of our notes.  
S.~Sint acknowledges support by the European Commission under
grant No.~FMBICT972442.


\begin{thebibliography}{99}

\bibitem{BardeenI}
W. Bardeen et al., Phys. Rev. D57 (1998) 1633

\bibitem{MP}
A. Hoferichter et al., Nucl. Phys. B (Proc. Suppl.) 63 (1998) 164

\bibitem{Schierholz}
M. G\"ockeler et al., Nucl. Phys. B (Proc. Suppl.) 73 (1999) 889
 
\bibitem{Aoki}
S. Aoki, Phys. Rev. D30 (1984) 2653

\bibitem{BardeenII}
W. Bardeen, A. Duncan, E. Eichten and H. Thacker,
Phys. Rev. D59 (1999) 014507

\bibitem{Hasenfratz}
P. Hasenfratz, Nucl. Phys. B525 (1998) 401 

\bibitem{LuscherI}
M. L\"uscher, Phys. Lett. B428 (1998) 342

\bibitem{GW} 
P.H. Ginsparg and K.G. Wilson, Phys. Rev. D25 (1982) 2649

\bibitem{Neuberger}
H. Neuberger, Phys. Lett. B417 (1998) 141 

\bibitem{LuscherII}
M. L\"uscher, Nucl. Phys. B549 (1999) 295

\bibitem{workinprogress}
R. Frezzotti, P.A. Grassi, S. Sint and P. Weisz,
work in progress

\bibitem{refl_pos} K. Osterwalder and E. Seiler, Ann. of Phys. 110
(1978) 440; G. Immirzi and K. Yoshida, Nucl. Phys. B210[FS6] (1982) 499 and
references therein
           
\bibitem{BochicchioEtAl}
M. Bochicchio et al., 
Nucl. Phys. B262 (1985) 331

\bibitem{paperIV}
M. L\"uscher, S. Sint, R. Sommer and H. Wittig, 
Nucl. Phys. B491 (1997) 344 

\bibitem{paperI}
M. L\"uscher, S. Sint, R. Sommer and P. Weisz,
Nucl. Phys. B478 (1996) 365

\bibitem{SW}
B. Sheikholeslami and R. Wohlert,
Nucl. Phys. B259 (1985) 572

\bibitem{paperIII}
M. L\"uscher et al., 
Nucl. Phys. B491 (1997) 323

\bibitem{Weinberg}
S. Weinberg, Phys. Rev. D8 (1973) 3497

\bibitem{letter}
K. Jansen et al.,
Phys. Lett. B372 (1996) 275

\bibitem{Stefan}
S. Sint, 
Nucl. Phys. B421 (1994) 135; Nucl. Phys. B451 (1995) 416

\bibitem{paperII}
M. L\"uscher and P. Weisz,
Nucl. Phys. B479 (1996) 429

\bibitem{PeterStefan}
S. Sint and P. Weisz, Nucl. Phys. B502 (1997) 251; 
Nucl. Phys. B (Proc. Suppl.) 63 (1998) 856

\bibitem{KarlRainer}
K. Jansen and R. Sommer, Nucl. Phys. B530 (1998) 185

\bibitem{PHMC}
R. Frezzotti and K. Jansen, \hbox{Phys. Lett.} B402 (1997) 328;
hep-lat/9808011;  hep-lat/9808038

\bibitem{Bitar}
K.M. Bitar, Phys. Rev. D56 (1997) 2736;  
Nucl. Phys. B (Proc. Suppl.) 63 (1998) 829 

\bibitem{bermionI}
S.J. Anthony, C.H. Llewellyn Smith and J.F. Wheather, Phys. Lett. 116B
(1982) 287

\bibitem{bermionII}
G.M. de Divitiis et al., Nucl. Phys. B455 (1995) 274 and references therein 

\bibitem{bermionIII}
R. Petronzio, Nucl. Phys. B (Proc. Suppl.) 42 (1995) 942

\bibitem{SharpeSingleton}
S. Sharpe and R. Singleton, Phys. Rev. D58 (1998) 074501

 \end{thebibliography}
\end{document}